\documentstyle[12pt]{article}

\begin{document}

\newcommand{\gsim}
{\mathrel{\raise.3ex\hbox{$>$\kern-.75em\lower1ex\hbox{$\sim$}}}}
\newcommand{\lesssim}
{\mathrel{\raise.3ex\hbox{$<$\kern-.75em\lower1ex\hbox{$\sim$}}}}

\newcommand{\beq}{\begin{equation}}
\newcommand{\eeq}{\end{equation}}
\newcommand{\bea}{\begin{eqnarray}}
\newcommand{\eea}{\end{eqnarray}}
\newcommand{\rv}{\rangle}
\newcommand{\lv}{\langle}

\begin{titlepage}

\begin{center}

{\Large \bf Large Lepton Number and High Temperature 
Symmetry Breaking in MSSM}

\vskip 1cm

Borut Bajc\footnote{borut.bajc@ijs.si}

{\it J. Stefan Institute, 1001  Ljubljana, Slovenia} 

\vskip 0.5cm

Goran Senjanovi\'c\footnote{goran@ictp.trieste.it}

{\it International Center for Theoretical Physics,
34100 Trieste, Italy }

\end{center}

\vskip 1cm

\centerline{\large \bf ABSTRACT}

\vskip 0.5cm
It is known that a large neutrino number, of the order of 
a few percent of the entropy of the universe, leads to 
symmetry breaking at high temperature. We show here that in 
the minimal supersymmetric standard model (MSSM) this implies 
the breaking of electromagnetic charge invariance at $T\gg M_W$ 
allowing for the solution of the monopole problem.

\end{titlepage}

\newpage

\section{Introduction}
\label{intr}

Although we know the baryon number of the universe with a 
fair precision, $B/S\approx 10^{-9\pm 1}$ ($S$ is the entropy 
of the universe), remarkably enough we know almost nothing about 
the lepton number. The only thing we can say with some confidence 
is the observational limit \cite{ks92,as98,riotto99,lp99} 

\begin{equation}
L\lesssim S\;.
\label{slexp}
\end{equation}

The determination of $L$ remains one of the main challenges for 
cosmology. Of course, it is just a part of what is probably 
{\it the} main missing link for the standard big-bang model 
of cosmology: the detection of the predicted neutrino sea. It 
is worth commenting that from the electric charge neutrality of 
the universe we know that the electron number is as small as the 
proton number. The lepton number of the universe, if not small, 
must be in the form of neutrinos.

Now, one often runs into an almost textbook argument (erroneous) 
that $|L|\approx |B|$, due to the anomaly of $B+L$. Namely, the 
$B+L$ violating processes at $T\gg M_W$ are presumably in equilibrium, 
so that $B+L$ gets washed out: $B+L\approx 0$. This is wrong, since the 
cosmology of $L$ large is rather subtle. It is known that in the standard 
model a large $L\gsim S/100$ leads to non-restoration of the SU(2)$_L$ 
gauge symmetry at $T\gg M_W$ \cite{l76,brs97}, and thus the sphaleron
processes remain exponentially suppressed \cite{ls94}, much as 
at $T=0$. More precisely, if $L\lesssim S/100$, then it 
is true that $B+L$ will be washed out and $L/S\approx 10^{-10}$, but 
in the opposite case one has a completely consistent cosmological 
scenario.

Thus, we should be open minded and allow for a possibility of 
$L/S$ not being small. The question, of course, is how natural 
is to have a substantial $L/S$. Well, in the context of supersymmetric 
theories, there is a simple mechanism of Affleck and Dine \cite{ad85} of 
producing large $L$ and $B$ through flat directions 
\cite{gelmini97}. The challenge, really, is to understand 
how come $B$ is so small. This important issue remains out 
of the scope of our work; instead, we just pursue the 
possibility that $L$ is large.

We are motivated, in large part, by a desire 
to solve the infamous mo\-no\-po\-le 
problem of grand unified theories. Monopoles get over produced during 
the phase transition in the early universe when one goes from the 
unbroken to the broken phase. If the symmetries are not restored 
at high $T$ \cite{w74,ms79}, then the non-restoration of GUTs will 
solve the monopole problem \cite{dms95}. Another alternative 
possibility is that U(1)$_{em}$ is broken above $T\approx M_W$
\cite{lp80}, 
so that the monopoles can be annihilated through the cosmic 
strings which interpolate between monopoles and anti-monopoles. 
Although it appears difficult to non-restore gauge symmetries in 
ordinary theories \cite{bl96}, in supersymmetry this may happen 
quite naturally through the presence of flat directions 
\cite{dk98,bs98}. The learned reader may know of the no-go 
theorem for symmetry non-restoration in supersymmetry \cite{h82,m84}. 
Although there has been some controversy in this issue \cite{dt96}, 
it is believed that without flat directions internal symmetries 
in SUSY are always restored \cite{bms96}.

On the other hand a large lepton number, as we mentioned before, 
provides a natural way of non-restoring symmetries. 
This has been recently addressed in the context 
of the minimal standard model, where one concludes 
that the Higgs sector must be enlarged, and 
unfortunately in an arbitrary fashion \cite{brs97}. 
Since the phenomenon of symmetry non-restoration 
with an external charge works as well in 
supersymmetry \cite{rs97}, it is natural to see what happens in the 
minimal supersymmetric standard model (MSSM). It turns out that 
the MSSM is tailor-fit for this task. Both the electromagnetic 
charge invariance and the SU(2) weak symmetry are necessarily 
broken at high $T$ for sufficiently large lepton asymmetry. 

Another interesting application of the large lepton number is 
in the ultra-high energy cosmic rays, as suggested in \cite{gk99}. 
Here the lepton number is used to help understand the cosmic rays 
with energies in excess of the Greisen-Zatsepin-Kuzmin cutoff. 

\section{MSSM at large $L$ and large $T$}
\label{mssm}

It is well known that the charged scalar field will 
have a non-vanishing vev at high temperature if the 
background charge of the universe is large enough. 
Take a simple example of a global U(1) symmetry with 
a field $\phi$ carrying a nonvanishing charge. From

\begin{equation}
V(\phi)=m^2|\phi|^2+\lambda|\phi|^4
\end{equation}

\noindent
one gets an effective potential at high $T$ ($\gg m$) 
and a charge density $n$ \cite{hw82,bbd91}

\begin{equation}
V_{eff}=-\mu^2|\phi|^2+{\lambda\over 3}T^2|\phi|^2+
\lambda|\phi|^4-\mu^2{T^2\over 6}+\mu n\;,
\end{equation}

\noindent
where $\mu$ is the chemical potential. It is clear that for 
$\mu>\sqrt{\lambda/3}T$, the U(1) symmetry will be necessarily 
broken even at $T\gg m$.

This simple fact suffices to establish that U(1)$_{em}$ would 
be broken in the MSSM for large enough background lepton number. 
Namely, both the left-handed slepton doublet $\tilde L$ and the 
right-handed selectron $\tilde e^c$ will get vevs; this implies 
the complete breaking of SU(2)$_L\times $U(1)$_Y$ gauge invariance.

The interesting property of this scenario is that not just 
the sleptons, but also the Higgses and the squarks $\tilde q$ 
learn about the lepton number through the gauge interactions 
and can in principle develop non-vanishing vevs. In order 
to solve the monopole problem, we need to make sure that for 
$M_W\ll T<M_X$, the unbroken group contains no U(1) subgroup. 

The question is if the possible breaking of SU(3)$_c$ through 
$<\tilde q>\ne 0$ can leave a U(1) subgroup, and the answer 
is no, as we show now. A first $<\tilde q>\ne 0$ breaks SU(3)$_c$ 
to its SU(2) subgroup (no U(1) factor remains) and the rest of 
the squarks are either doublets or singlets under this group. 
The doublets, if they get vevs, break SU(2) completely. 
Depending on who gets the vev (if any), SU(3)$_c$ 
is either preserved, or broken down to SU(2), or broken 
completely.

In short, whatever the original GUT symmetry is, at $T\gg M_W$ 
there is no U(1) factor and the monopoles, if at all created, 
necessarily annihilate through the cosmic strings attached to 
monopole-anti-monopole pairs.

Now, we wish to compute the critical lepton number density which 
suffices to have U(1)$_{em}$ broken for $T\gg M_W$. 

{\bf A toy model: leptons only.} {\hskip 1cm} To understand what is 
going on and for the sake of illustration we present first 
the one generation model of leptons only. This is truly a toy 
model for not only there are no quarks and Higgses, but as such the 
model is not anomaly free. We ignore this issue here since 
for us this is only a pedagogical illustration of what happens 
in the consistent and realistic theory discussed below. 

The $T=0$ potential contains only the D-terms

\begin{equation}
V^{(D)}={g'^2\over 2}(|\tilde e^c|^2-{1\over 2}|\tilde L|^2)^2+
{g^2\over 8}|\tilde L|^4\;.
\end{equation}

The high temperature correction to the effective potential 
can be obtained from \cite{ggs81}

\begin{equation}
\label{dvt}
\Delta V_T^{(D)}={g'^2 T^2\over 2}(|\tilde e^c|^2+{1\over 4}|\tilde L|^2) 
+{3 g^2 T^2\over 8}|\tilde L|^2\;.
\end{equation}

Clearly, the above terms $V^{(D)}$ and $\Delta V_T^{(D)}$ imply 
symmetry restoration at high temperature, a well known fact 
in supersymmetric theories.

Next, we need the corrections due to the nonzero lepton charge. 
This is usually done by introducing a chemical potential for 
each conserved charge \cite{brs97}, thus one for the lepton 
number ($\mu_L$) and one for the hypercharge ($\mu_Y$), which 
gives \cite{hw82,bbd91,hy74}

\begin{eqnarray}
\label{dvnmu0}
\Delta V_n=&-&(-\mu_L+\mu_Y)^2|\tilde e^c|^2-
(\mu_L-{\mu_Y\over 2})^2|\tilde L|^2\nonumber\\
&-&(-\mu_L+\mu_Y)^2{T^2\over 4}-
(\mu_L-{\mu_Y\over 2})^2{T^2\over 2}+\mu_Ln_L\;.
\end{eqnarray}

As one can see, the role of chemical potential is to provide 
negative mass square for the sleptons, which per se would imply 
symmetry breaking. Of course, whether or not the symmetry is broken, 
depends whether or not the chemical potential is bigger than the 
temperature.

Notice that we did not include the chemical potential for the weak 
isospin. The reason for this will be clear from what follows below. 
Also, we neglected the $\mu^4$ terms, which arise from the fermion 
loops; we checked however, that such an approximation is safe and 
changes the results only by a few percent. For this reason 
we will not include them in the rest of the paper.

Now it is more convenient to work directly with the lepton density 
and to eliminate chemical potentials. After all, the latter are 
derived quantities; the physical quantity is the lepton charge 
asymmetry. From the constraints \cite{hw82,bbd91}

\begin{equation}
{\partial \Delta V_n\over\partial\mu_i}=0\;\;,\;\; i=L,Y,
\end{equation}

\noindent
we get 

\begin{equation}
\Delta V_n={2n_L^2\over T^2+2|\tilde L|^2}
+{n_L^2\over T^2+4|\tilde e^c|^2}\;.
\end{equation}

Notice again that the charge density prefers nonvanishing vevs 
and the issue now is whether $n_L$ is bigger than $\approx T^3$ 
or not. The total effective potential 

\begin{equation}
V_{eff}=V^{(D)}+\Delta V_T^{(D)}+\Delta V_n
\end{equation}

\noindent
is minimized when $\partial V_{eff}/\partial|\phi|=0$
($\phi=\tilde e^c$, $\tilde L$), i.e. when 

\begin{eqnarray}
2|\tilde e^c|\left[g'^2\left(|\tilde e^c|^2-{|\tilde L|^2\over 2}\right)
+{g'^2\over 2}T^2-{4n_L^2\over (T^2+4|\tilde e^c|^2)^2}\right]=0\;,&&
\label{dvde}\\
2|\tilde L|\left[
{g^2\over 4}|\tilde L|^2
-{g'^2\over 2}\left(|\tilde e^c|^2-{|\tilde L|^2\over 2}\right)
+{g'^2+3g^2\over 8}T^2-
{4n_L^2\over (T^2+2|\tilde L|^2)^2}\right]=0\;.&&\label{dvdl}
\end{eqnarray}

The solution depends on the value of $n_L$. At 
small (compared with $T^3$) lepton charge density 
the vevs of $|\tilde e^c|$ and $|\tilde L|$ are trivially 
zero. Since $g'^2<(g'^2+3g^2)/4$ (we will use throughout the 
values $g'\approx 0.36$, $g\approx 0.65$) the first vev to become 
nonzero with growing $n_L$ is $|\tilde e^c|$. This happens for 
$n_L>n_L^0$, where 

\begin{equation}
\label{nl00}
n_L^0={g'\over 2\sqrt{2}}T^3\approx 0.13 \; T^3
\end{equation}

\noindent
is thus the charge density above which the U(1) symmetry is 
spontaneously broken. Naively one could think that this is 
enough to solve the monopole problem. However, as mentioned above, 
if the SU(2) symmetry is not broken, the sphalerons become 
operative and wash out any nonzero $B+L$, destroying our 
scenario. Thus the SU(2) symmetry must also be broken through 
a nonvanishing vev of $|\tilde L|$. This happens when the 
charge density is bigger than the critical value $n_L^c$ 
which is obtained by equating to zero the 
expressions in parenthesis in (\ref{dvde})-(\ref{dvdl}) for 
$|\tilde L|=0$. The equation for $n_L^c$ one gets is of the third 
order, whose analytic solution is not very illuminating. The 
numerical result is 

\begin{equation}
n_L^c \approx 0.20 \; T^3\;.
\label{nlc0}
\end{equation}

\noindent
In order to see whether such a value is experimentally 
allowed, one must calculate the entropy density at $T\gg M_W$, 
when our analysis make sense. Since in the early 
universe the entropy density $s\approx 100\; T^3$ 
for the MSSM, it is clear that the critical density 
(\ref{nlc0}) is well below the experimental 
limit $n_L\lesssim 100\; T^3$ from (\ref{slexp}).

The reader may now ask why we did not include a chemical 
potential also for the third component of the weak isospin. 
The point is, that it is zero as long as the symmetry is 
conserved, i.e. for all $n_L$ up to the critical charge density 
(\ref{nlc0}). This is different from the hypercharge chemical 
potential, entirely due to the non abelian character of the 
weak isospin. Since our aim was to calculate the critical 
charge density, we can assume from the beginning 
a vanishing weak isospin chemical potential. 

{\bf MSSM: one-generation case.} {\hskip 1cm} 
Let us now study in detail the complete case of 1-generation 
MSSM. Although the effect will be roughly the same as in the 
previous model, the exact values of the critical charges will 
turn out to differ up to a factor of 4. This surprisingly large 
number is unexpected: a closer inspection shows that it is due 
to the large number of degrees of freedom (especially in connection 
with the hypercharge) carried by the rest (mainly quarks superfields) 
of the MSSM spectrum. The fact confirm our findings in 
\cite{brs97}, i.e. that {\it all} the degrees of freedom must be 
taken into account, even those, such as quarks, that do not carry 
lepton number.

In order to simplify the computation we will consider the 
following special case.

a) The number of symmetries (chemical potentials) are 
taken as small as possible. Since the disappearance of 
Yukawa couplings brings new symmetries, we will study the MSSM 
effective potential for $T\lesssim 10^7$ GeV and for $tan\beta=1$. 
In this case all the Yukawa couplings are effective, i.e. 
their interactions $|y|^2 T$ are bigger than the Hubble 
constant $H\approx T^2/M_{Pl}$. Raising the temperature would make 
of course some Yukawas drop out of equilibrium, giving new 
conserved U(1) symmetries and thus new chemical potentials. 
Although straightforward, such a computation would not bring more 
insight into the problem, so we will limit ourselves to 
low enough temperatures (but of course still larger than $M_W$). 

b) In the MSSM with the vanishing soft terms and zero $\mu$ 
term in the superpotential there are two extra symmetries, 
one of which is an R-symmetry. 
In the above temperature regime neither of these symmetries 
is conserved: the interactions that violate them have the 
effective rate $\Gamma\approx\mu^2/T$, with 
$\mu\approx 10^2$ GeV. Clearly this is bigger than the 
expansion rate of the universe. 
This fact simplifies further the calculations, so that 
only three U(1) symmetries need to be considered: lepton number, 
baryon number and hypercharge. As in our previous toy model example,
chemical potentials connected to non abelian diagonal generators 
are zero as long as these symmetries are unbroken.

c) Now that we found the relevant symmetries involved, we can forget 
the small contributions in the superpotential (even the top Yukawa 
coupling turns out to be irrelevant) and take simply $W=0$. 

d) The 1-generation MSSM gives a very good estimate for the 
critical charge. We thus present our computations 
only for this case, but give at the end the numerical results 
for the realistic case of 3 generations.

\vskip 0.5cm

We have seen in the previous example that in order to determine 
the charge $n_L^0$, above which the first field gets 
a nonzero vev, one needs to consider only the quadratic 
(in fields) terms in the effective potential. 

The first of two different such 
terms is the leading 1-loop high temperature corrections to the 
scalar mass term (a special case is (\ref{dvt}))

\begin{equation}
\label{dvtdgen}
\Delta V_T^{(D)}={1\over 2}\sum_{a,i} C_2^a(r_i)g_a^2T^2|\phi_i|^2\;,
\end{equation}

\noindent
where $\phi_i$ are the fields in the representation 
$r_i$ of the gauge group, $a$ runs over different gauge 
groups with coupling constants $g_a$ and $C_2^a(r_i)=
(Y_i/2)^2$ for U(1) and $3/4$, $4/3$ for the fundamental 
representation of SU(2), SU(3) respectively. 
Written explicitly, (\ref{dvtdgen}) has the form 

\begin{eqnarray}
\label{dvtd}
\Delta V_T^{(D)}={T^2\over 2}&[&\left({g'^2\over 4}+
{3g^2\over 4}\right)\left(|H_u|^2+|H_d|^2+|\tilde L|^2\right)
+g'^2|\tilde e^c|^2\nonumber\\
&+&\left({g'^2\over 36}+{3g^2\over 4}+{4g_s^2\over 3}\right)
|\tilde Q|^2\nonumber\\
&+&\left({4g'^2\over 9}
+{4g_s^2\over 3}\right)
|\tilde u^c|^2+\left({g'^2\over 9}+{4g_s^2\over 3}\right)
|\tilde d^c|^2\;\;]\;.
\end{eqnarray}

The other term needed to determine $n_L^0$ is 
a generalization of (\ref{dvnmu0}). Using the rules 
developed in \cite{hw82,bbd91,hy74} one can 
work out a general expression for the contribution of 
the chemical potentials to the effective potential \cite{b98}

\begin{equation}
\label{dvnmu}
\Delta V_n=-{1\over 2}\mu_a{\cal M}_{ab}\mu_b+\mu_a n_a\;,
\end{equation}

\noindent
with 

\begin{equation}
\label{m}
{\cal M}_{ab}={T^2\over 6}\left(\sum_i f_a^i f_b^i + 
2\sum_i b_a^i b_b^i\right)+2\sum_i b_a^i b_b^i |\phi_i|^2\;.
\end{equation}

The indices $a$, $b$ run over the conserved symmetries ($\mu_a$ are 
the corresponding chemical potentials), while $i$ runs over the 
different fields; $f_a^i$ ($b_a^i$) is the $a$-th charge of the 
$i$-th fermion (boson) field; $\phi_i$ is the vev of the $i$-th 
boson field and $n_a$ is the value of the total $a$-th charge 
asymmetry (we will assume that only the lepton charge asymmetry 
is nonzero). We should stress that the formulae (\ref{dvnmu}) and 
(\ref{m}) are valid in any model, supersymmetric or not. Notice 
however, that we have neglected higher orders in 
$\mu$. The next term, $\mu^4$, will be shown to be much smaller 
than the terms in (\ref{dvnmu}) in our case.

Using the usual constraints

\begin{equation}
{\partial\Delta V_n\over\partial\mu_a}=0\rightarrow
\mu_a=({\cal M}^{-1})_{ab}n_b
\end{equation}

\noindent
(\ref{dvnmu}) becomes

\begin{equation}
\label{dvn}
\Delta V_n={1\over 2}n_a({\cal M}^{-1})_{ab} n_b\;.
\end{equation}

\noindent
If there is no R-symmetry, then $f_a^i=b_a^i$ ($=q_a^i$) and 

\begin{equation}
\label{mq}
{\cal M}_{ab}={T^2\over 2}\left(\sum_i q_a^i q_b^i \right)
+2\sum_i q_a^i q_b^i |\phi_i|^2\;.
\end{equation}

The meaning of (\ref{mq}) can be already obtained in the toy model 
(see (\ref{dvnmu0})). In that case the charges are $q_L=-1$, $1$, 
$q_Y=1$, $-1/2$ for the singlet and the doublet respectively. Of course, 
in such a simple model it is much more transparent not to use the 
complicated general formula (\ref{mq}). 

The analog of $n_L^0$ shown in (\ref{nl00}) is given by the minimal 
charge satisfying at least one of the equations

\begin{equation}
\label{dvdf}
{\partial V_{eff}\over\partial |\phi_i|^2}|_{|\phi_j|^2=0}=0\;,
\end{equation}

\noindent
where the relevant terms in $V_{eff}$ are given by 
(\ref{dvtdgen}) and (\ref{dvn}). To understand the meaning of 
this equation it may help to know that in the toy model 
discussed above (\ref{dvdf}) reduce to the expressions in the 
square brackets in (\ref{dvde})-(\ref{dvdl}) with vanishing vevs. 

If we write (\ref{m}) or (\ref{mq}) in a compact form 

\begin{equation}
{\cal M}_{ab}={T^2\over 2}{\cal A}_{ab}+\sum_i{\cal B}_{ab}^i|\phi_i|^2
\end{equation}

\noindent
and use the relation

\begin{equation}
{\partial\over\partial |\phi_i|^2}{\cal M}^{-1}|_{|\phi_j|^2=0}=
-\left({2\over T^2}\right)^2{\cal A}^{-1}{\cal B}^i{\cal A}^{-1}\;,
\end{equation}

\noindent
we get

\begin{equation}
n_L^0=\min_i\left[{T^6\left(\sum_aC_2^a(r_i)g_a^2\right)\over
4\left({\cal A}^{-1}{\cal B}^i{\cal A}^{-1}\right)_{LL}}\right]^{1/2}\;.
\end{equation}

Using the fields' charges of Table \ref{tab1} it is now straightforward 
to calculate the matrices ${\cal A}$ and ${\cal B}$ and find out that 
$n_L^0$ is given by the value, when the right-handed selectron 
(just as in the toy model) begins to get a nonzero vev:

\begin{equation}
\label{nl0}
n_L^0=\sqrt{2}g'T^3\approx 0.50\; T^3\;.
\end{equation}

For $n_L>n_L^0$, U(1)$_Y$ is spontaneously broken by 
$|\tilde e^c|^2\ne 0$. Notice that $n_L^0$ in the 1-generation MSSM 
(\ref{nl0}) is exactly 4 times bigger than the one found in the 
simplified model with only leptons (\ref{nl00}). This may be somewhat 
of a surprise. After all, quarks and Higgs superfields carry no lepton 
charge, so why should their presence matter so much? 
The fact that they do matter was already noticed 
in our previous work on the standard model \cite{brs97}, but here 
the effect is even more dramatic. The point is that any particle with 
non-zero hypercharge learns about the lepton number through its interactions 
with the photon and the $Z$ boson. 

\begin{table}[h]
\begin{center}
\begin{tabular}{|c|c|c|c|c|c|c|c|c|}\hline
${\cal Q}$ & $e^c$ &  $L$ & $H_d$ & $H_u$ & $Q$ & $u^c$ & $d^c$ &  
name \\ \hline
1          &  $-1$ &   1  &   0   &   0   &  0  &   0   &   0   & 
$L$ \\ \hline
2          &    0  &   0  &   0   &   0   & 1/3 &$-1/3$ &$-1/3$ & 
$B$ \\ \hline
3          &    1  &$-1/2$&$-1/2$ &  1/2  & 1/6 &$-2/3$ &  1/3  & 
$Y/2$ \\ \hline
\end{tabular}
\end{center}
\caption{\label{tab1} The relevant charges $q_i^a$ ($a=L,B,Y/2$, while 
$i$ goes over the chiral superfields) in MSSM.}
\end{table}

To get the critical charge $n_L^c$, where also SU(2)$_L$ gets 
broken presumably by $|\tilde L|^2$, one must take into account that now 
$|\tilde e^c|^2\ne 0$. For this reason the full dependence of $V_{eff}$ on 
$|\tilde e^c|^2$ must be retained for $n_L>n_L^0$, but for $n_L<n_L^c$ only 
the quadratic $|\phi_i|^2$ term needs to be considered for all the 
remaining scalar fields (as was the case for all the fields when 
$n_L<n_L^0$).

From the general D-term 

\begin{equation}
\label{vdgen}
V^{(D)}={1\over 2}\sum_aD^aD^a\;\;,\;\;
D^a=-g_a\phi_i^\dagger (T^a)_{ij}\phi_j\;,
\end{equation}

\noindent
with $(T^a)_{ij}$ the generators of the group in the matrix 
representation (equal to $Y/2$ for U(1)$_Y$ and $\tau^a/2$,
$\lambda^a/2$ for the fundamental representation of SU(2)$_L$, 
SU(3)$_c$ respectively) the following D-terms from (\ref{vdgen}) 
are relevant for the determination of the critical charge

\begin{equation}
\label{vd}
{g'^2\over 2}|\tilde e^c|^4+g'^2|\tilde e^c|^2 
\left(-{1\over 2}|\tilde L|^2-{1\over 2}|H_d|^2+{1\over 2}|H_u|^2
+{1\over 6}|\tilde Q|^2-{2\over 3}|\tilde u^c|^2+
{1\over 3}|\tilde d^c|^2\right)\;.
\end{equation}

The critical charge is the minimal charge which satisfies 
at least two equations among (\ref{dvdf}), one of which is for 
$i=|\tilde e^c|$ and $|\tilde e^c|^2\ne 0$ now. One has thus two 
equations for the two unknowns, $|\tilde e^c|^2$ and $n_L^c$: first 

\begin{equation}
\label{eqec}
{\partial V_{eff}\over\partial |\tilde e^c|^2}(|\tilde e^c|^2,n_L^c)=0
\end{equation}

\noindent
and second, the equation among

\begin{equation}
\label{eqphi}
{\partial V_{eff}\over\partial |\phi_i|^2}(|\tilde e^c|^2,n_L^c)=0
\end{equation}

\noindent
for $i\ne |\tilde e^c|$ with the smallest solution for $n_L^c$. Now, 
if you see a huge shark in your immediate vicinity while swimming 
far away from the shore, what do you do? A correct answer to this 
question gets you a free coffee from either of the authors. After 
having established that you are following this stuff, let us go on.

The effective potential is found from (\ref{dvtd}), (\ref{dvn}) and 
(\ref{vd})

\begin{equation}
V_{eff}=V^{(D)}+\Delta V_T^{(D)}+\Delta V_n\;.
\end{equation}

Using the notation

\begin{equation}
V_T=V^{(D)}+\Delta V_T^{(D)}\;,
\end{equation}

\noindent
one can write down the equations (\ref{eqec}), (\ref{eqphi}) 
more explicitly as 

\begin{equation}
{\partial V_T\over\partial |\phi_i|^2}(|\tilde e^c|^2)=
{(n_L^c)^2\over 2}\left[
\left({T^2\over 2}{\cal A}+{\cal B}^{|\tilde e^c|}|\tilde e^c|^2\right)^{-1}
{\cal B}^i
\left({T^2\over 2}{\cal A}+{\cal B}^{|\tilde e^c|}|\tilde e^c|^2\right)^{-1}
\right]_{11}\;.
\end{equation}

It is now straightforward to find out numerically that the first 
equation to be satisfied among (\ref{eqphi}) is for $i=|\tilde L|$, 
as expected, which breaks the SU(2)$_L$ gauge symmetry. The value 
of the critical charge comes out to be 

\begin{equation}
n_L^c\approx 0.58\; T^3\;,
\end{equation}

\noindent
which is almost 3 times bigger than in the toy model (\ref{nlc0}). 
For $n_L>n_L^c$ both SU(2)$_L$ and U(1)$_Y$ are broken, solving thus 
the monopole problem.

At this point one can also check the numerical value of 
the $\mu^4$ terms neglected in (\ref{dvnmu}). For example from 
\cite{hy74} one finds that for each fermion the ratio between 
the $\mu^4$ term and $\mu^2$ term is $[\mu/(\pi T)]^2$, which 
turns out to be always smaller than $0.02$ for $n_L\le n_L^c$, 
which we considered above. This is somewhat different from a similar 
calculation in the non-supersymmetric standard model \cite{brs97}, 
where a larger charge, which is needed to break the SU(2) symmetry, 
leads to larger chemical potentials and thus the quartic terms 
turn out to be still important (corrections to the critical charge 
in SM are sometimes of order $50\%$).

Furthermore, the inclusion of the Yukawa couplings does 
not change the 
numerical results for the critical charge: in fact, 
the value of the lepton Yukawa coupling is far too small to 
contribute considerably, while the top Yukawa does not enter 
the relevant equations, since the quarks and the Higgses do not 
get a nonzero vev at this stage (we have checked explicitly this 
statement confirming the above results). Obviously, since 
the Yukawa interaction is unimportant, it is irrelevant which 
generation is to be considered in this one-generation MSSM.

{\bf The three-generations realistic case.} {\hskip 1cm} 
The last generalization to be done is to include all three 
generations. This we did numerically along the lines 
described above. We assumed that the lepton charge is 
stored in only one of the three generations (remember that there 
are three conserved lepton number symmetries for vanishing or 
very small neutrino masses). The more complicated case with 
more than one non-vanishing lepton asymmetry can be treated along 
the same lines as here, but in our opinion it would teach us 
nothing new. Furthermore, one expects no significant change 
in our results (we will see below that going from one to three 
generations makes very little difference). The shark is still 
here: what are you doing about it? 

The results are given in Table (\ref{tab2}) 
together with the previously obtained critical charges. The 
relative error between the results from the 1-generation 
and 3-generations MSSM are of the order of $20\%$, which is 
not much, comparable to other errors we can expect (higher loop 
contributions, etc.). We can thus say, that the much simpler 
1-generation MSSM gives a very good estimate of the correct 
result. This result confirms the approximate value for the 
critical charge which leads to the breaking of the SU(2)$_L$ symmetry 
in the non-supersymmetric standard model \cite{brs97}.
Recall, however, that in the non-supersymmetric version the 
electromagnetic invariance is not broken and thus the monopole 
problem is not solved. Thus supersymmetry is seen to play an 
essential role in our considerations. Without it one would have to 
postulate an {\it ad hoc} existence of light charged scalar 
particles if one wants the electromagnetic invariance 
spontaneously broken at high temperature.

\begin{table}[h]
\begin{center}
\begin{tabular}{|c|c|c|c|c|c|c|c|c|}\hline
            & toy model & 1-gen. MSSM & 3-gen. MSSM \\ \hline
$n_L^0/T^3$ &   0.13    &    0.50     &    0.43     \\ \hline
$n_L^c/T^3$ &   0.20    &    0.58     &    0.72     \\ \hline
\end{tabular}
\end{center}
\caption{\label{tab2} The charges, above which the U(1) ($n_L^0$) 
and SU(2) ($n_L^c$) symmetries are broken, from the models 
described in the text.}
\end{table}

\section{Discussion and outlook}

The phenomenon of symmetry breaking at high temperature plays 
an important role in cosmology. First, it provides a simple 
way-out of the monopole problem in grand unification. Furthermore, 
it changes completely the conventional scenario of baryogenesis, 
since it implies a freezing of sphaleron effects at all 
temperatures. A particularly simple realization of this 
mechanism is provided by the large lepton number of the 
universe. In this paper we have shown how a sufficiently 
large lepton number implies the complete breaking of 
SU(2)$\times$U(1) symmetry in the MSSM for $T\gg M_W$. This is 
achieved through nonzero VEVs of at least one left-handed 
and one right-handed sleptons. 

The critical value for the lepton density of the universe 
is in complete agreement with the observation. We hope that 
our work provides a further impetus for the improvement of the 
existing experimental limits. On the other hand, we need 
a more fundamental explanation for the large lepton number 
and the small baryon number of the universe. We hope that 
this is only a matter of time, or, that we will learn that 
the lepton number of the universe is small too. 
It should be mentioned, though, that even a small lepton 
number would not necessarily imply that our scenario is 
completely wrong. In fact, in order to have the SU(2)$_L\times$
U(1)$_Y$ symmetry completely broken, one does not need to have 
necessarily a large lepton number: what is important is to have 
a large conserved charge density, whatever it is \cite{md99}. 
For example, a large $R$ charge \cite{brs98} could do the job as well. 
In such a scenario the monopole problem would be automatically 
solved, but we could not test it, since the $R$ charge would 
have been washed out at smaller temperatures, when the 
supersymmetry (and thus $R$ symmetry) breaking terms become 
important. This is why the existence of a large lepton 
number for solving the monopole problem is much more 
interesting. It is testable, at least in principle. 

This mechanism of symmetry non-restoration could maybe also 
provide a way-out of the false vacuum problem of supersymmetric 
GUTs \cite{w82}. However, here the issue is somewhat subtle and must 
be addressed separately. The point is that it is not enough to keep 
the SU(5) unbroken above $T\approx M_X$; one must also show that the 
vev of the adjoint {\bf 24} Higgs is non-zero and 
connects correctly to its physical value at small temperature. 

We should also mention that symmetry non-restoration at high 
temperature may solve the domain wall problem of spontaneously 
broken discrete symmetries \cite{ds95}. Now, in the MSSM this is 
not the case, but in many extensions of the theory one may have 
discrete symmetries. For example it can be spontaneous CP violation 
or a $Z_3$ \cite{asw95} discrete symmetry of the next-to-minimal
supersymmetric standard model with the singlet field instead of 
a $\mu$ term in the superpotential. Clearly, the large lepton 
number is tailor-fit also for this task. 

Another interesting question can be raised regarding the lepton 
number and neutrino masses. Namely, the solar and the 
atmospheric neutrino data point strongly towards nonvanishing 
neutrino masses \cite{s99}. If neutrino mass is of 
Dirac form, then of course it has no impact on the lepton 
number of the universe. If on the other hand neutrino mass 
is of Majorana nature, as many of us suspect and as suggested 
by the see-saw mechanism \cite{seesaw}, then the situation 
becomes problematic. After all, Majorana masses break lepton 
number and one would expect any background lepton number 
to be washed out during the expansion of the universe. The 
issue however is more subtle and the outcome depends on 
the size of neutrino mass and the temperature range of one's 
interest \cite{lss82}. 

\section*{Acknowledgements}

We wish to acknowledge useful discussions with Charan Aulakh, 
Umberto Cotti and Alejandra Melfo. 
The work of B.B. was supported by the  Ministry 
of Science and Technology of Slovenia and that of 
G.S. in part by EEC under the TMR contract 
ERBFMRX-CT960090. B.B. thanks ICTP for hospitality, 
where part of this work was done.

\end{document}